\documentclass[twocolumn, prl, a4paper, 10pt, superscriptaddress, nopacs, floatfix]{revtex4-1}
\usepackage{siunitx}
\usepackage{graphicx}
\usepackage{xcolor}
\usepackage{physics}
\usepackage{nicefrac}
\usepackage{bm}
\usepackage{float}
\usepackage[utf8]{inputenc}
\usepackage[colorlinks, citecolor=blue, linkcolor=black]{hyperref}
\usepackage{ulem}
\usepackage{booktabs}
\usepackage{multirow}
\usepackage{xspace}

\newcommand{\VB}{V$_\text{B}^-$\xspace}

\begin{document}

\title{Optically-active spin defects in few-layer thick hexagonal boron nitride}

\author{A. Durand}
\thanks{Contributed equally to this work.}
\author{T. Clua-Provost}
\thanks{Contributed equally to this work.}
\author{F. Fabre}
\author{P. Kumar}
\affiliation{Laboratoire Charles Coulomb, Université de Montpellier and CNRS, 34095 Montpellier, France}
\author{J. Li}
\author{J.~H.~Edgar}
\affiliation{Tim Taylor Department of Chemical Engineering, Kansas State University, Kansas 66506, USA}
\author{P.~Udvarhelyi}
\affiliation{Department of Atomic Physics, Budapest University of Technology and Economics, H-1111 Budapest, Hungary}
\author{A. Gali}
\affiliation{Department of Atomic Physics, Budapest University of Technology and Economics, H-1111 Budapest, Hungary}
\affiliation{Wigner Research Centre for Physics, P.O. Box 49, H-1525 Budapest, Hungary}
\author{X.~Marie}
\author{C.~Robert}
\affiliation{Université de Toulouse, INSA-CNRS-UPS, LPCNO, 135 Avenue Rangueil, 31077 Toulouse, France}
\author{J. M. Gérard}
\affiliation{Université Grenoble Alpes, CEA, Grenoble INP, IRIG, PHELIQS, Grenoble, France}
\author{B. Gil}
\author{G. Cassabois}
\author{V. Jacques}
\email{vincent.jacques@umontpellier.fr}
\affiliation{Laboratoire Charles Coulomb, Université de Montpellier and CNRS, 34095 Montpellier, France}

\begin{abstract}
Optically-active spin defects in hexagonal boron nitride (hBN) are promising quantum systems for the design of two-dimensional quantum sensing units offering optimal proximity to the sample being probed. In this work, we first demonstrate that the electron spin resonance frequencies of boron vacancy centres (\VB) can be detected optically in the limit of few-atomic-layer thick hBN flakes despite the nanoscale proximity of the crystal surface that often leads to a degradation of the stability of solid-state spin defects. We then analyze the variations of the electronic spin properties of \VB centres with the hBN thickness with a focus on (i) the zero-field splitting parameters, (ii) the optically-induced spin polarization rate and (iii) the longitudinal spin relaxation time. This work provides important insights into the properties of \VB centres embedded in ultrathin hBN flakes, which are valuable for future developments of foil-based quantum sensing technologies.
\end{abstract} 
\date{\today}

\maketitle

Spin defects with optically detectable magnetic resonances in hexagonal boron nitride (hBN) are currently attracting a deep scientific interest for the deployment of quantum sensing technologies on a two-dimensional (2D) material platform~\cite{RevModPhys.89.035002,tetienne2021}. Among several optically-active spin defects recently discovered in hBN~\cite{Aharanovich_NatMat2021,Chejanovsky2021,Stern2022,Guo2021,Gottscholl2020}, the negatively-charged boron vacancy (V$_{\rm B}^-$) centre stands out due to its well-established atomic structure~\cite{Gottscholl2020,ivady2020,haykal2021decoherence} and ease of creation by various irradiation methods~\cite{IgorImplant2020,ACSomega2022,PhysRevB.98.155203,LaserWriting2021,nano11061373,Li2021}. This defect features a spin triplet ground state whose electron spin resonance (ESR) frequencies can be interrogated by optical means~\cite{Gottscholl2020} and strongly depends on external perturbations such as magnetic fields, strain, and temperature~\cite{GottschollNatCom2021,ACSPhot_Guo2021,GaoStrain2022,IgorStrain2022}. Such properties make the V$_{\rm B}^-$ centre in hBN a promising candidate for the design of a flexible 2D quantum sensing unit, which could offer an ultimate atomic-scale proximity between the spin-based sensor and the probed sample.\\
\indent In this context, recent proof-of-concept experiments have demonstrated that ensembles of \VB centres can be employed for magnetic imaging in van der Waals heterostructures~\cite{Tetienne2023,Du2021,PhysRevApplied.18.L061002}. However, the hBN flakes used in these works were a few tens of nanometers thick, {\it i.e.} far from the 2D limit. A pending question towards future developments of hBN-based quantum sensing foils is whether \VB centres retain their spin-dependent optical response in atomically-thin hBN layers. In this Letter, we give a positive answer to this question. We first show that the ESR frequencies of \VB centres remain optically detectable in the limit of few-atomic-layer thick hBN flakes
despite the nanoscale proximity of the crystal surface that often leads to a degradation of the stability of solid-state spin defects~\cite{PhysRevB.82.115449,PhysRevB.83.081304}. We then study how the electronic spin properties of \VB centres evolve with the hBN thickness with a focus on (i) the zero-field splitting parameters, (ii) the optically-induced spin polarization rate and (iii) the longitudinal spin relaxation time.\\
\indent As starting material, we employ a millimeter-sized hBN crystal isotopically purified with $^{10}$B~\cite{Liu2018}, in which V$_{\rm B}^-$ centres are created by neutron irradiation [see Methods]. The thermal neutron capture cross-section of $^{10}$B being one of the largest of the periodic table~\cite{Cataldo2017}, neutron irradiation of a monoisotopic h$^{10}$BN crystal ensures an efficient creation of V$_{\rm B}^-$ centres through nuclear transmutation doping~\cite{haykal2021decoherence,Li2021}. hBN flakes are obtained by mechanical exfoliation of this neutron-irradiated crystal and then transferred on a SiO\textsubscript{2}($90$~nm)/Si substrate. The thickness of the resulting hBN flakes is measured by atomic force microscopy (AFM), while the optical and spin properties of \VB centres are studied with a scanning confocal microscope employing a green laser excitation, a high numerical aperture objective (NA=0.8), a photon counting module and an external loop antenna for microwave excitation. Unless otherwise stated, experiments are performed under ambient conditions.\\
 \begin{figure*}[t!]
  \centering
  \includegraphics[width = 18cm]{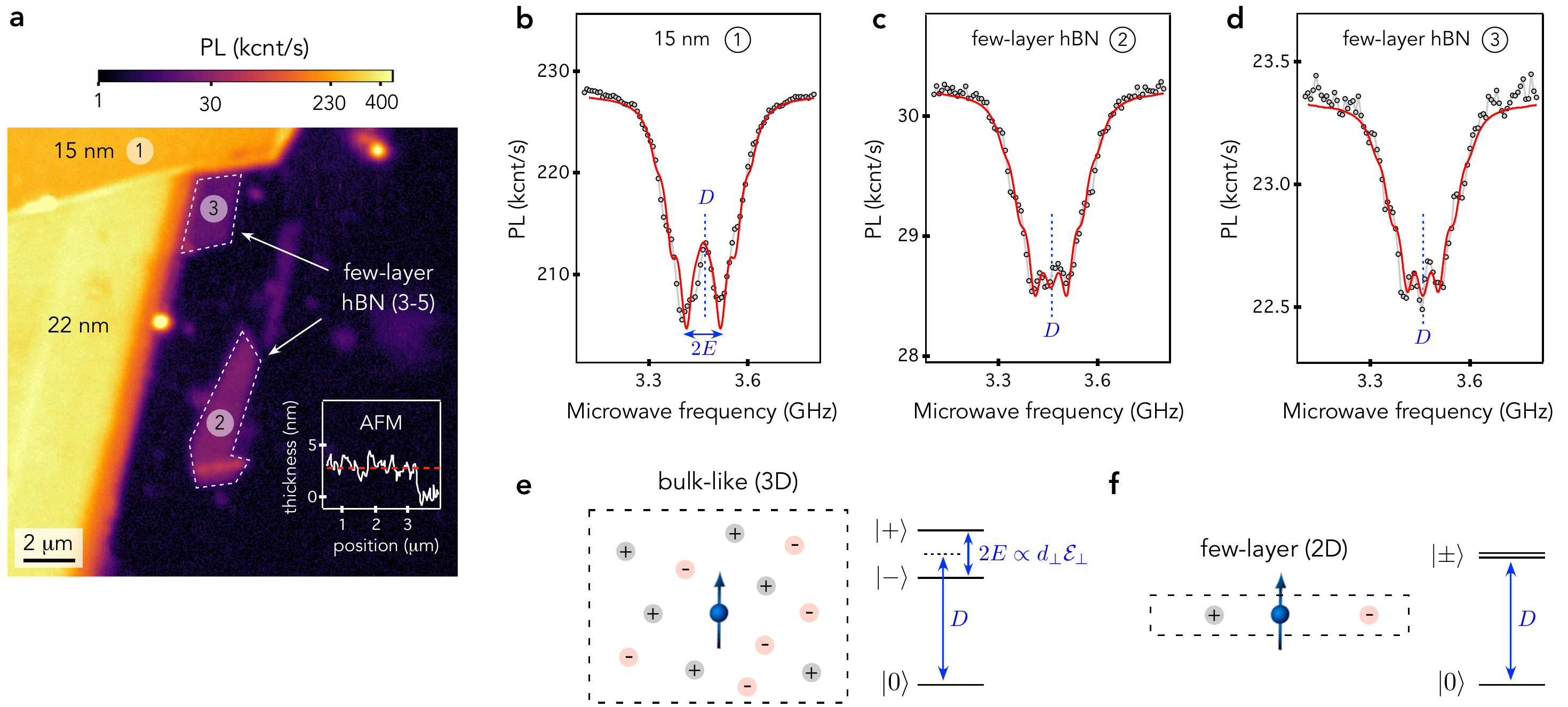}
  \caption{(a) PL raster scan of hBN flakes of different thicknesses doped with \VB centres. Inset: line-profile of the AFM topography recorded across the flake labeled \raisebox{.5pt}{\textcircled{\raisebox{-.9pt} {2}}}, corresponding to few-layer hBN. (b-d) Optically-detected ESR spectra recorded at zero external magnetic field (b) for the 15-nm-thick hBN flake labeled \raisebox{.5pt}{\textcircled{\raisebox{-.9pt} {1}}} and (c,d) for few-layer hBN  labeled \raisebox{.5pt}{\textcircled{\raisebox{-.9pt} {2}}} and \raisebox{.5pt}{\textcircled{\raisebox{-.9pt} {3}}}. The decreased ESR contrast in few-layer hBN results from a reduced signal to background ratio. (e-f) Sketch showing a central \VB electronic spin (blue arrow) surrounded by charged defects and the resulting energy level structure of its triplet ground state (e) for a bulk-like hBN flake and (f) for few-layer hBN. In panels (b-d), the red solid line shows the result of a simulation considering the interaction of the \VB electronic spin with an electric field produced by surrounding charged defects (see Methods).}
  \label{fig2}
\end{figure*}
\indent Photoluminescence (PL) raster scans show that all exfoliated hBN flakes produce a PL signal whose spectrum features a broad emission line centered around $800$~nm that corresponds to the characteristic emission of \VB centres~\cite{Gottscholl2020}. For the thinnest flakes obtained in this work, AFM measurements indicate a typical thickness around $3$~nm [see line-profile in Fig.~\ref{fig2}(a)]. Although the interlayer distance in bulk hBN is $0.33$~nm, AFM images of monolayers deposited on SiO$_2$ usually lead to a characteristic height that can be as large as $2$~nm owing to the presence of a thin contamination or water layer between the substrate and hBN~\cite{Gorbachev2011,Rousseau2021}. We therefore estimate that our thinnest flakes correspond to few-layer ($\sim 3$ to $5$) hBN. The PL image shown in Fig.~\ref{fig2}(a) reveals that \VB centres remain easily detectable in such ultrathin hBN flakes, albeit with a reduced signal to background ratio.\\
\indent We focus on the electronic spin properties of \VB centres and their variations with the hBN thickness. The \VB centre has a spin triplet ground state with an axial zero-field splitting $D\sim 3.46$~GHz between a singlet state $|m_s=0\rangle$ and a doublet $|m_s=\pm1\rangle$, where $m_s$ denotes the spin projection along the $c$-axis of the hBN crystal. Under optical illumination, spin-selective processes provide an efficient polarization of the $\rm{V}_{\rm B}^{-}$ centre in state $|m_s=0\rangle$. In addition, the PL response is stronger when the state $|m_s=0\rangle$ is predominantly populated. These two properties enable to record ESR spectra by sweeping the frequency of a microwave excitation while monitoring the spin-dependent PL signal~\cite{Gottscholl2020}. When the microwave frequency is resonant with a transition between the electron spin sublevels of the \VB centre, a drop of the PL signal is detected.  A typical ESR spectrum recorded at zero external magnetic field on a 15-nm-thick hBN flake (bulk-like) is shown in Fig.~\ref{fig2}(b) as reference. We detect the two characteristic magnetic resonances of the \VB spin triplet ground state, whose frequencies are given by $\nu_{\pm}=D\pm E$ with $D=3.46$~GHz and $E=61$~MHz. These values are identical to those commonly obtained for \VB centres embedded in bulk hBN crystals~\cite{Gottscholl2020,haykal2021decoherence}. Importantly, ESR spectra can also be recorded for few-layer thick hBN, showing that \VB centres retain their magneto-optical properties in the 2D limit [Fig.~\ref{fig2}(c-d)]. However, while the axial zero-field splitting $D$ remains unchanged for few-layer thick hBN flakes, we observe a strong reduction of the $E$-splitting so that a single magnetic resonance is detected, in which the characteristic hyperfine structure of the \VB centre can even be resolved.\\
\indent The origin of the $E$-splitting has been often attributed to local strain effects in the hBN lattice. However, recent {\it ab initio} calculations of the strain susceptibility parameters have shown that the spin-strain interaction mainly induces fluctuations of the zero-field splitting parameter $D$~\cite{udvarhelyi2023planar}. Averaging over an ensemble of \VB centres undergoing different local deformations of the crystal should therefore lead to a single broadened magnetic resonance, in contrast with the experimental data obtained for bulk-like hBN flakes [Fig.~\ref{fig2}(b)]. It was recently suggested that the $E$-splitting rather results from the interaction of the \VB electronic spin with a local electric field~\cite{udvarhelyi2023planar,ChongZu2022}, as for the nitrogen-vacancy (NV) defect in diamond~\cite{PhysRevLett.121.246402}. Each negatively-charged \VB centre is likely accompanied by a positively-charged defect in order to ensure charge neutrality of the hBN crystal. These charges produce a local electric field, whose component perpendicular to the $c$-axis ($\mathcal{E}_\perp$) mixes the $|m_s=\pm1\rangle$ spin sublevels, leading to new eigenstates $| \pm \rangle$ separated by an energy $2E\propto d_{\perp}\mathcal{E}_\perp$, where $d_{\perp}$ is the susceptibility of the \VB centre to a transverse electric field [Fig.~\ref{fig2}(e)]. Moreover, since the coupling of the \VB centre to the electric field component along the $c$-axis is cancelled out by symmetry ($d_\parallel=0$), an electric field only induces a splitting of the spin states without modifying the parameter $D$~\cite{udvarhelyi2023planar}. Ensemble averaging then naturally leads to a zero-field ESR spectrum featuring two magnetic resonances, as commonly observed for \VB centres hosted in bulk-like hBN flakes.\\
\indent To support this qualitative discussion, we simulate the ESR spectrum by using a microscopic charge model originally introduced for NV defects in diamond~\cite{PhysRevLett.121.246402} and recently applied to \VB centres in hBN~\cite{udvarhelyi2023planar,ChongZu2022}. Positive and negative charges of equal density $\rho_c$ are randomly positioned around a central \VB electronic spin [Fig.~\ref{fig2}(e)]. We first calculate the transverse electric field at the \VB location, from which the splitting of the ESR frequencies is obtained by using $d_\perp=21$~Hz/(V.cm$^{-1}$), a value recently inferred from {\it ab initio} calculations~\cite{udvarhelyi2023planar}. Moreover, we include the hyperfine interaction of the \VB centers with the three neighboring $^{14}$N nuclear spins with a characteristic coupling constant $\mathcal{A}_{zz}=47$~MHz. The ESR spectrum is then simulated (i) by applying a convolution with a Gaussian profile to take into account the broadening of the \VB electron spin transitions and (ii) by averaging over a large ensemble of charge state configurations (see~Methods). This procedure is repeated while adjusting the charge density $\rho_c$ and the optically-detected ESR contrast to fit the experimental results. A good agreement between the simulation and the ESR spectrum recorded on a 15-nm-thick hBN flake is obtained for $\rho_c=0.054(5)$~nm$^{-3}$ [red solid line in Fig.~\ref{fig2}(b)].\\ 
\indent For such a charge density, our simulation indicates that the transverse electric field is mostly produced by the $\sim 10$ closest charges, which are localized within a sphere with diameter $d_c\sim 7$~nm. When the thickness $t$ of the hBN flake becomes smaller than $d_c$, the number of charges producing a sizable transverse electric field at each \VB site decreases by a factor $\sim d_c/t$. For hBN flakes close to the monolayer limit, the $E$-splitting then becomes too small to be detected [Fig.~\ref{fig2}(f)], thus leading to a single magnetic resonance in ESR spectra, as observed in our experiments. This effect, together with the appearance of a well-defined hyperfine structure, are both reproduced by the microscopic charge model applied to a monolayer hBN [red solid lines in Fig.~\ref{fig2}(c-d)]. The vanishing $E$-splitting is a first modification of the \VB electronic spin properties in the 2D limit. 
\begin{figure}[t]
  \centering
  \includegraphics[width = 8cm]{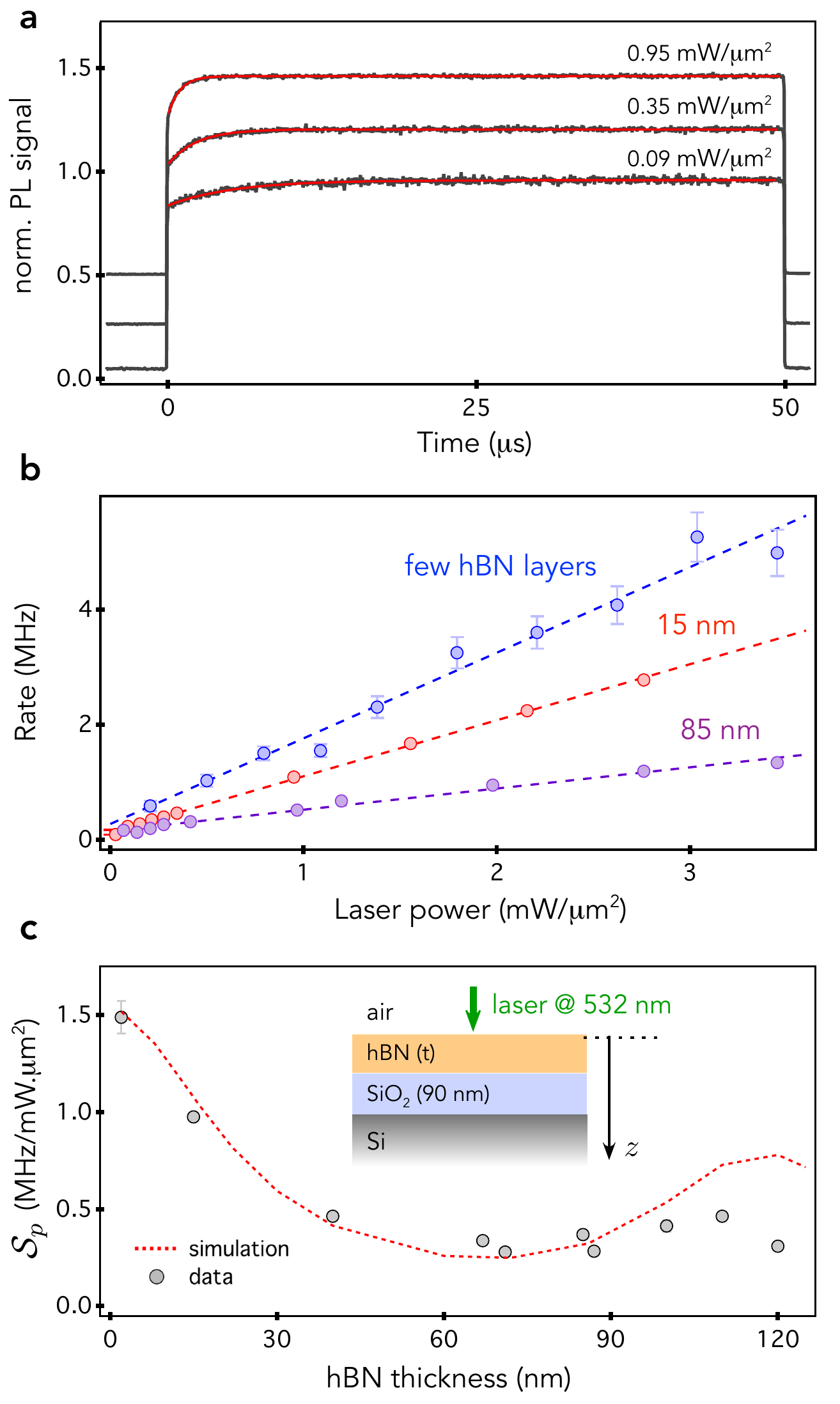}
  \caption{(a) PL time traces measured for \VB centres hosted in a $15$-nm-thick hBN flake during a $50$-$\mu$s-long laser pulse at three different laser powers. The data are vertically shifted for clarity and the red solid lines are data fitting with an exponential function.  (b) Spin polarisation rate $\mathcal{R}_p$ as a function of the laser power $\mathcal{P}$ for hBN flakes with different thicknesses. The dashed lines are linear fits from which the slope $\mathcal{S}_p=d\mathcal{R}_p/d\mathcal{P}$ is extracted. (c) Parameter $\mathcal{S}_p$ as a function of the hBN thickness. The red dashed line correspond to a simulation of the thickness-dependent absorption of \VB centres for a $532$~nm excitation wavelength.}
  \label{fig3}
\end{figure}

We then study the electron spin polarization rate under optical pumping through time-resolved PL measurements. Starting with the electronic spin in a thermal state - {\it i.e.} with equal populations in the ground state spin sublevels of the \VB centre - the PL signal is recorded during a $50$-$\mu$s-long laser pulse. As illustrated in Fig.~\ref{fig3}(a), the PL signal first increases at the beginning of the pulse before reaching a steady-state for which the \VB centres are polarized in state $|m_s=0\rangle$. Data fitting with an exponential function is used to infer the optically-induced polarization rate $\mathcal{R}_p$. In our experiments, the optical pumping power $\mathcal{P}$ is well-below the saturation of the \VB centre's optical transition. In this case, the polarization rate increases linearly with $\mathcal{P}$ [Fig.~\ref{fig3}(b)], with a slope $\mathcal{S}_p=d\mathcal{R}_p/d\mathcal{P}$ proportional to the absorption of \VB centres at the optical illumination wavelength. Interestingly, this slope is maximal for few-layer hBN [Fig.~\ref{fig3}(b,c)].\\ 
\indent To understand the thickness dependence of the spin polarization rate, the optical absorption is computed from the time-averaged Poynting vector describing the direction and magnitude of the electromagnetic energy flux at any depth $z$ of the sample~\cite{Yar}. Within a transfer matrix approach, the averaged Poynting vector is calculated, at the 532 nm-excitation wavelength, every $\delta z=2$~nm inside the air/hBN($t$)/SiO$_2$($90$~nm)/Si multilayer system [Fig.~\ref{fig3}(c)]. The presence of \VB centres is modeled by a complex linear susceptibility which is resonant with the excitation laser. For any hBN flake of thickness $t$, we thus obtain the spatially-resolved transmittance $T(z,t)$, from the ratio of the Poynting vector at a depth $z$ with the incident Poynting vector. The absorption per unit length $\alpha(z,t)$ is then given by $\alpha(z,t)\delta z=T(z,t)-T(z-\delta z,t)$. Remarkably, $\alpha(z,t)$ reaches its maximum close to the air/hBN interface ($z=0^+$) for ultrathin hBN flakes and for any thickness $t$ multiple of $\lambda/2n$, of the order of $120$~nm for a refractive index $n\sim2.3$ at $532$~nm~\cite{PhysRevMaterials.2.024001}. In Fig.~\ref{fig3}(c), each experimental point corresponds to the spin polarization rate averaged over the \VB centres distributed over the whole thickness of the hBN flake. We thus compare our data to the spatially-averaged value of the absorption per unit length $\overline{\alpha}(t)=1/t\times \int_0^t\alpha(z,t)dz$. An excellent agreement with our experimental data is reached, demonstrating the importance of photonic effects in the understanding of the spin properties of \VB centres in multilayer hBN. In particular, because $\alpha(z,t)$ is maximum for ultrathin flakes without any averaging, $\overline{\alpha}(t)\sim\alpha(z=0^+)$, there is a strong enhancement of the spin polarization rate in the limit of atomically-thin hBN.

\begin{figure}[t]
  \centering
  \includegraphics[width = 8.5cm]{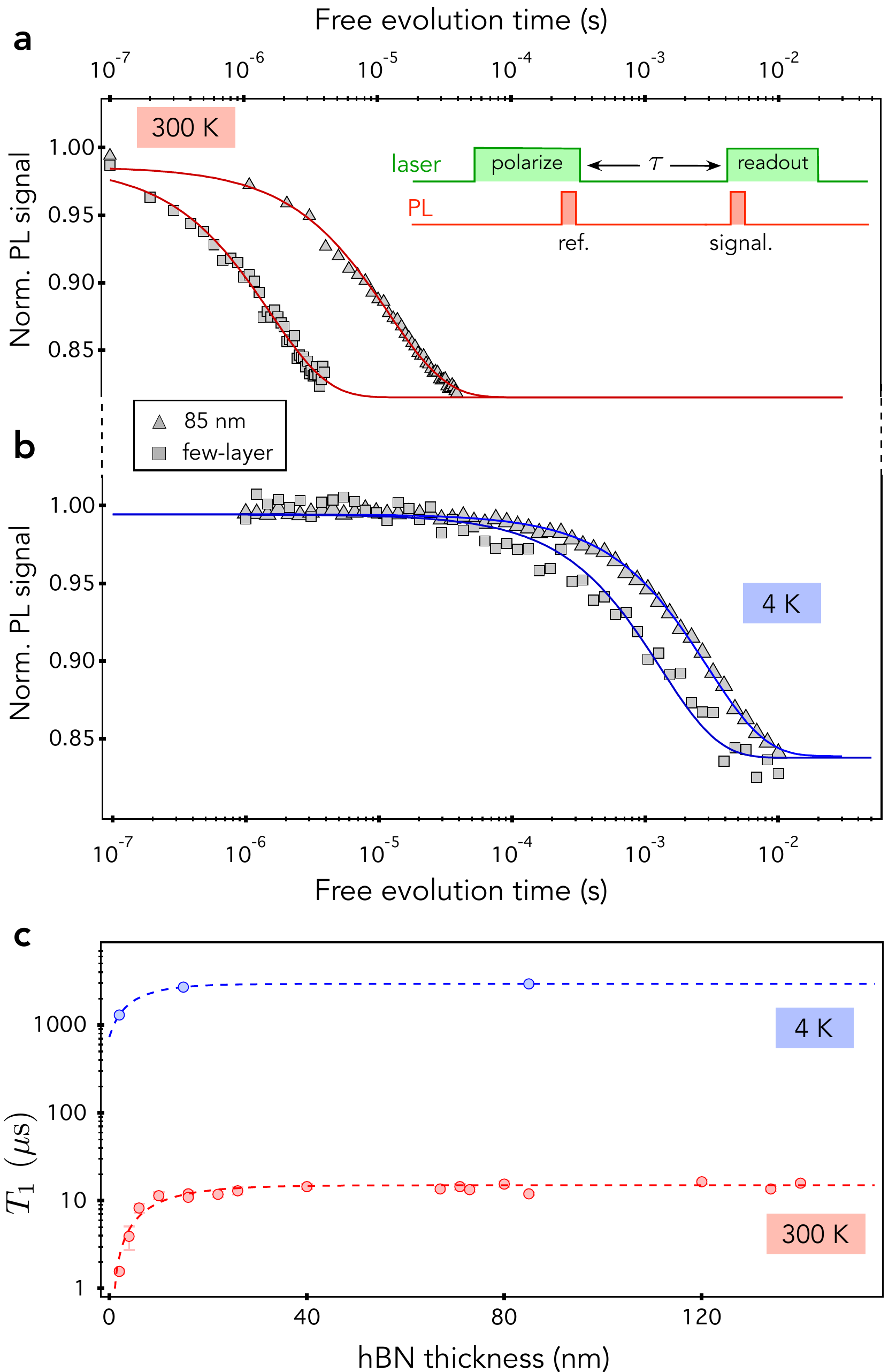}
  \caption{(a) Spin relaxation curves recorded at room temperature for \VB centres hosted in a 85-nm-thick hBN flake (triangles) and in few-layer-thick hBN (squares). The inset shows the experimental pulse sequence. The spin-dependent PL signal is integrated at the beginning of the readout pulse ($500$~ns window) and normalized using a reference PL value obtained at the end of the first laser pulse. (b) Measurements performed on the same hBN flakes at cryogenic temperature (4K). In (a) and (b), the solid lines are data fitting with an exponential decay. (c) Evolution of the $T_1$ time with the hBN thickness at room temperature (red points) and at 4K (blue points). The dashed lines are guides for the eye.}
  \label{fig4}
\end{figure}

As last experiments, we perform measurements of the \VB longitudinal spin relaxation time ($T_1$) by using the experimental sequence sketched in Fig.~\ref{fig4}(a). A laser pulse is first used to polarize the \VB centres in state $|m_s=0\rangle$ by optical pumping. After relaxation in the dark during a variable time $\tau$, the remaining population in $|m_s=0\rangle$ is probed by integrating the spin-dependent PL signal produced at the beginning of a readout laser pulse. The longitudinal spin relaxation time $T_1$ is then obtained by fitting the decay of the integrated PL signal with an exponential function. Typical spin relaxation curves recorded at room temperature for a bulk-like hBN flake and for few-layer hBN are shown in Fig.~\ref{fig4}(a). For the thick layer, we obtain $T_1\sim 13 \ \mu$s, a value similar to that usually observed for \VB centres hosted in bulk hBN crystals~\cite{haykal2021decoherence}. In this case, it was shown that spin relaxation is dominated by spin-phonon interactions~\cite{Gottscholl2021,Lunghi2022}. For few-layer hBN, the spin relaxation time drops to $T_1\sim 1 \ \mu$s [Fig.~\ref{fig4}(a)]. A systematic thickness-dependent study shows that $T_1$ starts to decrease for thicknesses below $\sim 10$~nm [Fig.~\ref{fig4}(c)]. \\
\indent The spin relaxation time of solid-state spin defects is commonly shortened when they are placed near the host crystal surface~\cite{PhysRevB.76.245306,PhysRevB.87.235436,PhysRevLett.112.147602,PhysRevX.9.031052}. This effect is often explained by considering that the interface is covered by a bath of randomly-fluctuating paramagnetic impurities. These fluctuations produce a magnetic noise that opens an additional channel for spin relaxation with a rate $T_{1,{\rm n}}^{-1}$. The relaxation time can then be expressed as 
\begin{equation}
T_1^{-1} = T_{1,{\rm p}}^{-1} + T_{1,{\rm n}}^{-1} \ ,
\end{equation}
where $T_{1,{\rm p}}$ is the contribution of spin-phonon interactions. Using this simple framework, the shortening of $T_1$ in ultrathin hBN flakes - {\it i.e.} for which the \VB centres are closest to the surface - can thus be explained by considering that spin relaxation becomes limited by surface-related magnetic noise. Dipolar magnetic coupling between surface paramagnetic impurities is one possible source of magnetic noise~\cite{PhysRevB.87.235436}. If this process dominates, the spin relaxation time of \VB centres hosted in ultrathin hBN flakes should not be improved at low temperature because intrabath dipolar coupling is temperature-independent~\cite{PhysRevB.102.165427}. To check this hypothesis, the experiments are reproduced in a cryogenic environment (4K). For bulk-like hBN flakes, $T_1$ increases up to few millisecond in agreement with a relaxation  driven by spin-phonon interactions. Strikingly, $T_1$ also reaches the millisecond range for few-layer hBN [Fig.~\ref{fig4}(b)]. We conclude that spin relaxation is not limited by a magnetic noise produced by dipolar coupling between surface-related paramagnetic impurities.\\
\indent Several phenomena could potentially explain both the shortening of the $T_1$ time in ultrathin hBN flakes at room temperature and its increase by three orders of magnitude in a cryogenic environment. First, a thermally-activated magnetic noise - {\it e.g.} produced by phonon-mediated fluctuations of surface paramagnetic impurities - could account for these experimental findings~\cite{PhysRevB.76.245306,PhysRevB.102.165427}. Alternatively, one may consider that the spin-phonon interaction of \VB centres is modified when the hBN crystal reaches the 2D limit. Nonlinear coupling terms with the lattice vibrations stem from anharmonicity. Recent time-resolved measurements of the nonlinear phonon response have evidenced an enhancement of anharmonic effects in ultrathin hBN~\cite{PhysRevB.104.L140302}, suggesting that the nonlinear spin-phonon coupling may also be increased for \VB centres in few-layer hBN. With the current set of experimental results, we cannot discriminate between these tentative explanations. 

To conclude, we have demonstrated that \VB centres retain their magneto-optical response in the limit of few-atomic-layer thick hBN flakes albeit with some modifications of their electron spin properties. The transition from a 3D host crystal to a 2D layer is first characterized by a strong reduction of the transverse zero-field splitting parameter $E$. This effect is well reproduced by considering the interaction of \VB centres with a local electric field whose amplitude vanishes in the 2D limit owing to a decreased number of surrounding charges. More generally, this result confirms that the transverse $E$-splitting of \VB centres is not related to strain effects in the hBN lattice. We then showed that the optically-induced spin polarization rate increases for few-layer hBN flakes owing to an enhanced absorption of \VB centres, which highlights the importance of photonic effects in thin hBN layers. We finally studied the variation of the \VB longitudinal spin relaxation with the hBN thickness. Although the $T_1$ time is reduced at room temperature in the 2D limit, it increases by three orders of magnitude in a cryogenic environment, reaching the millisecond range. All together, these results provide important insights into the properties of \VB centres embedded in ultrathin hBN flakes, which are valuable for future developments of foil-based quantum sensing technologies.

\section{Methods}
\noindent{\bf Neutron irradiated hBN crystal.} We rely on a bulk h$^{10}$BN crystal synthesized through the metal flux growth method described in Ref.~\cite{Liu2018}, while using a boron powder isotopically enriched with $^{10}$B ($99.2\%$). This  crystal has a typical lateral size in the millimeter range and a thickness of a few tens of micrometers. Neutron irradiation was performed at the Ohio State University Research Reactor, which produces a thermal neutron flux of $10^{12}$ neutrons$\cdot$cm$^{-2}\cdot$s$^{-1}$. The h$^{10}$BN crystal was exposed for 2 h and 25 min leading to a total fluence of $2.6 \times 10^{16}$ neutrons$\cdot$cm$^{-2}$. Neutron irradiation creates V$_{\rm B}^-$ centers through damages induced by neutron scattering through the crystal and via neutron absorption leading to nuclear transmutation~\cite{Li2021}. The optical and spin properties of \VB centres in this neutron-irradiated h$^{10}$BN crystal are described in Ref.~\cite{haykal2021decoherence}.\\

\noindent{\bf Simulation of ESR spectra.} In our microscopic charge model, we randomly place elementary point charges at the atomic sites of hBN within a simulation sphere with $10$~nm radius around a central V$_{\rm B}^-$ electron spin. The number of charges in the simulation sphere corresponds to an average charge density $\rho_c$. For a given charge distribution, we first calculate the electric field produced at the center of the sphere and then solve the spin Hamiltonian of the V$_{\rm B}^-$ center including the zero-field splitting, hyperfine and electric field interactions as described in details in Ref.~\cite{udvarhelyi2023planar}. The resulting ESR frequencies are then calculated from the diagonalized spin Hamiltonian written in the basis of the $|m_s\rangle$ spin sublevels of the triplet ground state. This procedure is reproduced for $10^4$ different random charge configurations in order to simulate an ensemble of V$_{\rm B}^-$ centres surrounded by different local charge distributions. The ESR spectrum is finally obtained by making the sum of all individual transitions convolved with a Lorentzian profile with a fixed width of $40$~MHz, in order to take into account the inhomogeneous broadening of the \VB electron spin transitions. We then optimise the charge density $\rho_c$ and the optically-detected ESR contrast to fit the experimental data using the least squares method. To simulate ESR spectra in the monolayer limit, we only take into account the charges placed in the layer containing the V$_{\rm B}^-$ centre.
 
The result of the fitting procedure is shown as red solid lines in Fig.~1(b)-(d) of the main text. For the 15-nm-thick hBN flake (bulk-like), we obtain a charge density $\rho_c=0.054(5)$~nm$^{-3}$ [Fig.~1(b)]. The standard deviation, which mainly originates from the asymmetry of the experimental ESR spectrum, is estimated from the statistics over $15$ simulation runs. We note that the extracted charge density is in good agreement with the one inferred for the bulk neutron-irradiated crystal used for exfoliation (see sample S2 in Ref.~\cite{udvarhelyi2023planar}).

For the ultrathin hBN flakes labeled \raisebox{.5pt}{\textcircled{\raisebox{-.9pt} {2}}} and \raisebox{.5pt}{\textcircled{\raisebox{-.9pt} {3}}} in the main text, the ESR spectra are simulated with a monolayer model, as explained above. The fitting procedure leads to $\rho_c=0.081(7)$~nm$^{-3}$ and $\rho_c=0.079(6)$~nm$^{-3}$ for flake \raisebox{.5pt}{\textcircled{\raisebox{-.9pt} {2}}} and \raisebox{.5pt}{\textcircled{\raisebox{-.9pt} {3}}}, respectively. The discrepancy between these values and that obtained for the bulk-like flake can be explained by the fact that our ultrathin hBN flakes are not monolayers. We note that the fitting procedure could also be done by fixing the charge density to the bulk value while using the number of layers as a fitting parameter. This approach would assume that the charge density remains constant when the thickness of the hBN flake decreases. However, in the limit of ultrathin flakes, additional charges located at the hBN surface could start to play an important role, making the analysis very difficult. Although not fully quantitative, our analysis confirms that the vanishing $E$-splitting in ultrathin flakes results from the reduction of charges surrounding the V$_{\rm B}^-$ centre.\\

\noindent {\it Acknowledgements} - This work was supported by the French Agence Nationale de la Recherche under the program ESR/EquipEx+ (grant number ANR-21-ESRE-0025), the Institute for Quantum Technologies in Occitanie through the project BONIQs and Qfoil, an Office of Naval Research award number N000142012474 and by the U.S. Department of Energy, Office of Nuclear Energy under DOE Idaho Operations Office Contract DE-AC07-051D14517 as part of a Nuclear Science User Facilities experiment. We acknowledge the support of The Ohio State University Nuclear Reactor Laboratory and the assistance of Susan M. White, Lei Raymond Cao, Andrew Kauffman, and Kevin Herminghuysen for the irradiation services provided.

\end{document}